# FEEDBACK CONTROL OF A QUANTUM DOT

C. M. MARCUS
*Department of Physics, Stanford University*
*Stanford, CA 94305-4060*

ABSTRACT

Mesoscopic structures are generally operated in an open-loop configuration, whereas most practical electronics including quantum interference devices such as SQUIDs are operated closed-loop, taking advantage of feedback. This paper presents some basic considerations on the use of feedback in mesoscopic samples with universal statistical properties. The controllability of mesoscopic fluctuations is shown to be connected to problems in continuum percolation, leading to the requirement of two control parameters to achieve robust control.

## 1. Introduction

This paper is an attempt to connect two active areas of research. One is mesoscopic physics, which is the study of quantum effects in structures sufficiently small that quantum coherence is important (typically ~1 μm scale or less) but sufficiently large that concepts of disorder and chaos are applicable.[1,2,3] The other is the control of quantum systems, which is the attempt to guide quantum processes, for instance by applying properly shaped optical pulses.[4,5,6]

The system we investigate is the quantum dot, a small island of charge connected to large electronic reservoirs by narrow (quantizing) leads. Quantum dots are man-made quantum systems: commonly, the size and shape of the device are defined using electrostatic gates patterned on the surface of a semiconductor heterostructure using electron-beam lithography, with minimum feature sizes around 40 nm. An example of such a dot is shown in Fig. 1.

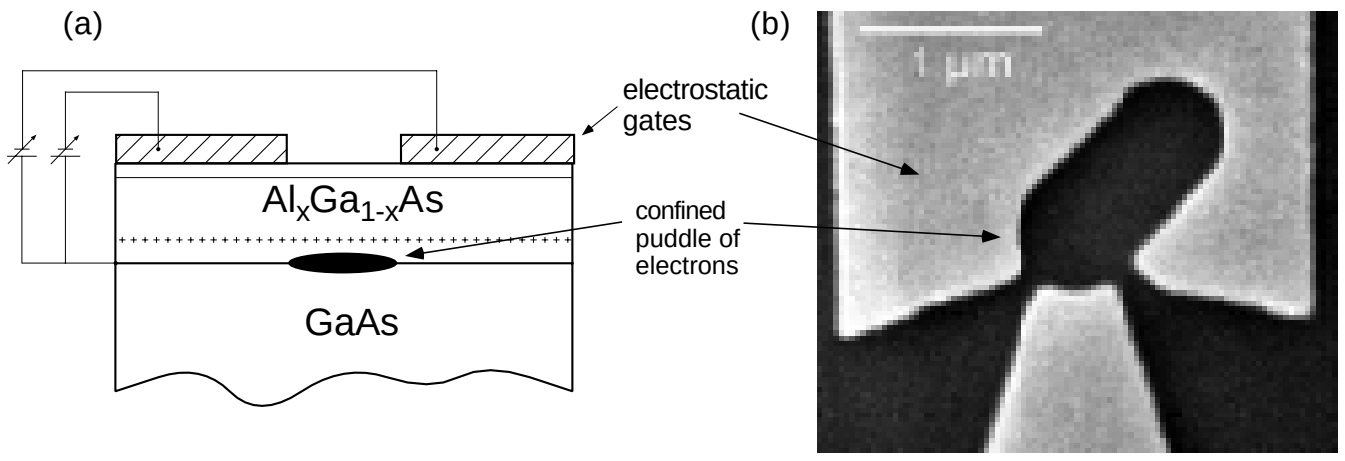

**Fig. 1**. Schematic side view and SEM micrograph top view of "stadium" quantum dot from Ref. 25. Electrostatic gates on the surface of a $GaAs/Al_xGa_{1-x}As$ (x ~ 0.07 is the Al concentration) heterostructure confine the electrons, allowing passage into the dot only through the narrow leads.

At low temperatures, in the range ~ 0.1K to 4K, a variety of quantum effects appear, including the quantization of charge in dots coupled via tunneling leads,[7,8] quantum interference effects such as conductance fluctuations and weak-localization in dots with open leads,[9,10,11] quantization of angular momentum reflected in shells structure in cylindrically symmetric dots,[12] and the Kondo effect in extremely small dots.[13] Rather than survey this rich field of research, we focus on a rather narrow problem concerning the use of feedback, and attempt to provide ample references as a guide to the literature. We have tried to select references intended for a broader scientific audience at the expense of providing an accurate record of exactly who did what first.

The main theme of this paper is that the universal statistical properties of quantum systems that are chaotic in their classical limit allows general statements to be made concerning the controllability of feedback systems based on mesoscopic quantum dots. This idea takes advantage of the fact that devices that are disordered or have shapes that generate chaotic dynamics (as illustrated in Fig. 2) possess universal statistical properties. Interestingly, the present results show an unexpected mapping from problems in feedback and control of mesoscopic systems onto problems in continuum percolation theory and statistical topography.[14]

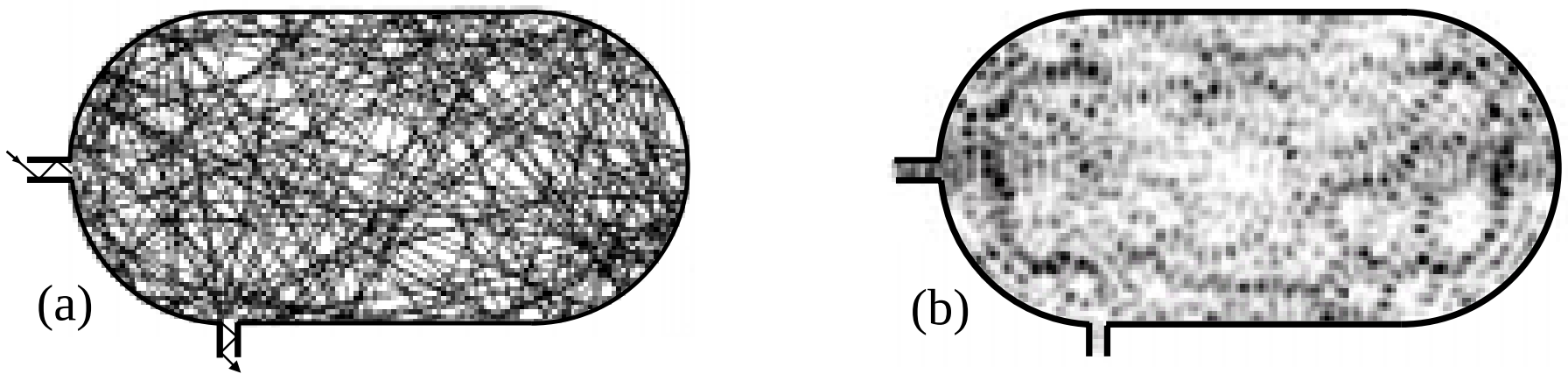

**Fig. 2**. (a) Chaotic classical trajectory entering, then bouncing within, and finally exiting an open stadium billiard. Signatures of *classical* chaos appear in the quantum transport as universal fluctuation statistics (b) A quantum wave function for the open stadium billiard (computed by R. Akis and D. Ferry), suggests an alternative approach more applicable to nearly-isolated devices, in which transport is described by the coupling of the dot wave function to the electron states in the leads.

The use of feedback in quantum systems is already well developed in the case of superconducting quantum interference devices, or SQUIDs. The present discussion could well be extended in directions already worked out for SQUIDs, for instance issues of optimization of design and operating parameters.[15] Some results in this direction are reported in Ref. 16. The important difference between quantum dots and SQUIDs is that the fluctuations due to quantum interference are random in the case of quantum dots, and thus amena-ble to a statistical treatment, whereas interference effects are regular (periodic in magnetic field $B$) for a SQUID. It is worth emphasizing that the results for quantum dots should apply to any nonsymmetric coherent conductor connected via two leads to bulk reservoirs.

## 2. Coherent Transport through Quantum Dots and Universal Statistics

At sufficiently low temperatures, electrons passing through a mesoscopic device maintain quantum coherence.[1,17] As a result, for disordered or chaotic structures, random but repeatable fluctuations in conductance are seen as external parameters such as a magnetic field or confining gate voltage are changed. A typical example of this effect is shown in Fig. 3, where the interesting combination of random fluctuations along with near-perfect symmetry about $B = 0$ is evident. The statistics and correlations of these fluctuations have been investigated theoretically using random matrix theory (RMT)[11] and nonlinear-sigma-model techniques[18], including the effects of decoherence[19,20] and temperature[21], and are generally in good agreement with experiment.[22]

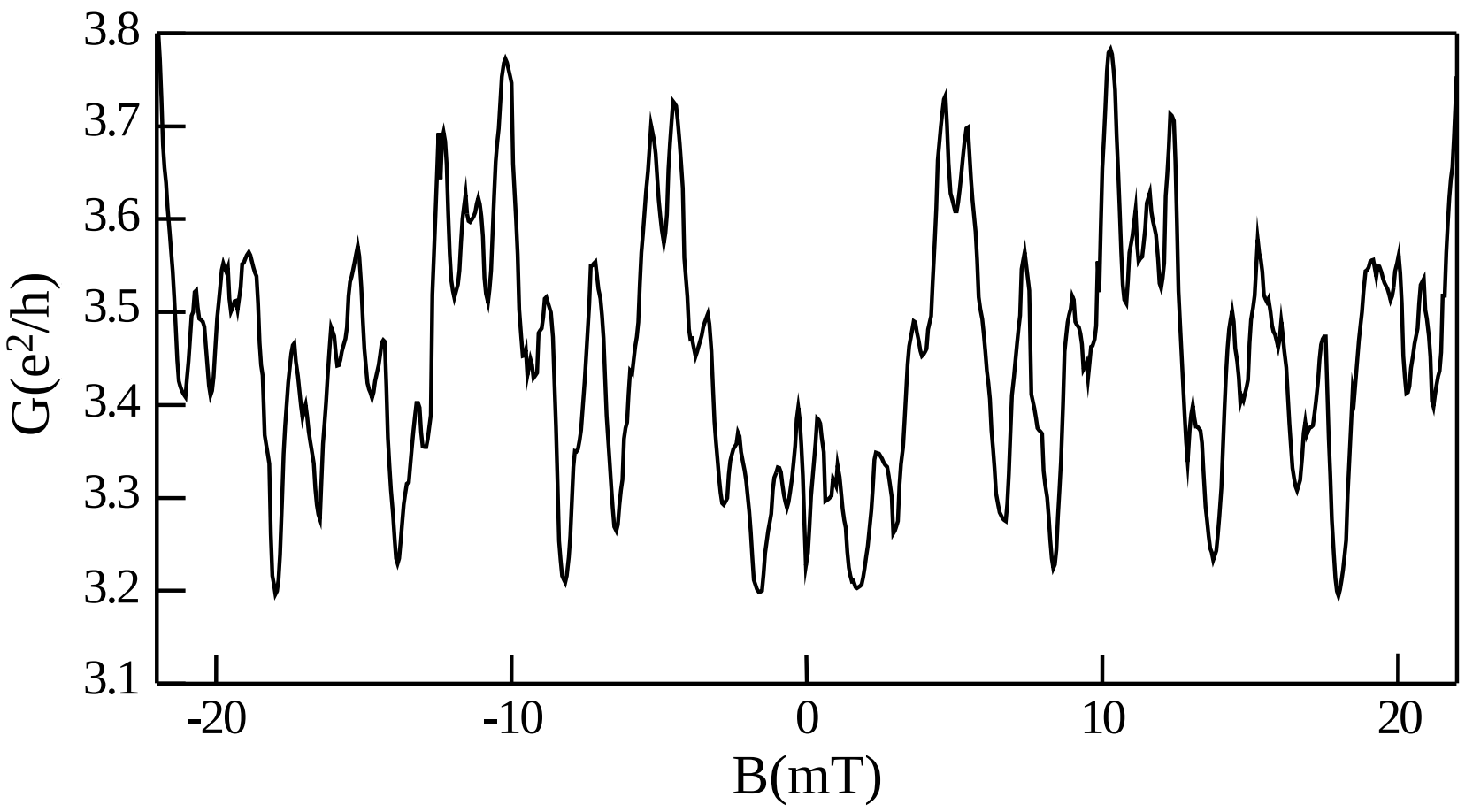

**Fig. 3.** A typical trace showing conductance fluctuations in a chaotic quantum dot at temperature $T \sim 100$mK. Notice the irregular pattern of conductance, but the nearly perfect symmetry about $B = 0$. The vertical scale is in units of the quantum of conductance, $e^2/h \sim 1/(25.8\text{k}\Omega)$

We summarize the relevant theoretical results: Fluctuations in dot conductance as external parameters are changed result from changes in the interference pattern of electrons flowing through the dot.[23,24,25] Resulting conductance fluctuations are statistically random with a nongaussian distribution for few modes coupling the dot to the reservoirs (the number of quantum modes correspond to the number of half wavelengths of the electron fitting laterally into the leads, where the electron wavelength is ~ 40 nm in GaAs, the material used for these experiments). Many modes coupling the dot to the reservoirs as well as finite temperature and dephasing all tend to make the distribution nearly gaussian, allowing the randomly fluctuating conductance to be fully characterized by its mean and variance. In this case, the RMT results for the mean and variance are as follows:

The **mean conductance** for a chaotic dot with two equivalent leads is just the familiar resistors-in-series expression for the two lead resistances,

$$\langle G_{dot} \rangle = \frac{2e^2}{h}\left(\frac{N}{2}\right) \tag{1}$$

where $N$ is the number of modes in each lead and $e^2/h$ ~ $(25.8\ \mathrm{k}\Omega)^{-1}$ is the quantum of conductance. Equation (1) applies except at $B = 0$ where pairs of backscattered trajectories following time-reversed paths always interfere constructively, causing a lower average conductance. This decrease in average conductance $\delta G \equiv \langle G_{dot} \rangle_{B\neq 0} - \langle G_{dot} \rangle_{B=0}$, is itself a quantum interference effect that depends on the dephasing rate,

$$\delta G = \left(e^2/h\right)\left[N/(2N+\gamma_\varphi)\right], \tag{2}$$

where $\gamma_\varphi = h/(\Delta\tau_\varphi)$ is the dimensionless dephasing rate, $\Delta$ is the mean quantum level spacing of the dot, and $\tau_\varphi$ is the coherence time. The dependence of $\delta G$ on dephasing provides a method for measuring $\tau_\varphi(T)$ in quantum dots.[17]

The **variance of conductance** depends on temperature $T$ as well as dephasing rate. For $N$ =1,

$$\mathrm{var}(G_{dot}) = \left(\frac{e^2}{h}\right)^2 \frac{\Delta}{6k_BT} \frac{1+\gamma_\varphi/2}{\left(a+b\gamma_\varphi\right)^2}, \tag{3}$$

where $a = \sqrt{3}\left(\sqrt{45/16}\right)$ and $b = 1\left(\sqrt{1/3}\right)$ for B≠0 (B=0).

Correlations in conductance as a function of magnetic field obey a Lorentzian-squared distribution at low temperatures,[23] $T << \Delta$, and become Lorentzian at higher temperatures.[21] The characteristic magnetic field $B_c$ corresponds to ~ one quantum of flux (~4 $mT\ \mu m^2$) through a typical trajectory passing through the dot, and is roughly set by the area of the dot, $B_c \sim \phi_o / A_{dot}$. The correlation function in gate voltage is Lorentzian,[26] with a char-acteristic voltage scale, $V_g$. A typical landscape of conductance fluctuations as a function of magnetic field and shape distorting gate voltage is shown in Fig. 4. Good agreement between theory and experiment for the full distributions of conductance is demonstrated in Fig. 5, where experimental distributions were sampled over a variety of shape configurations.[22]

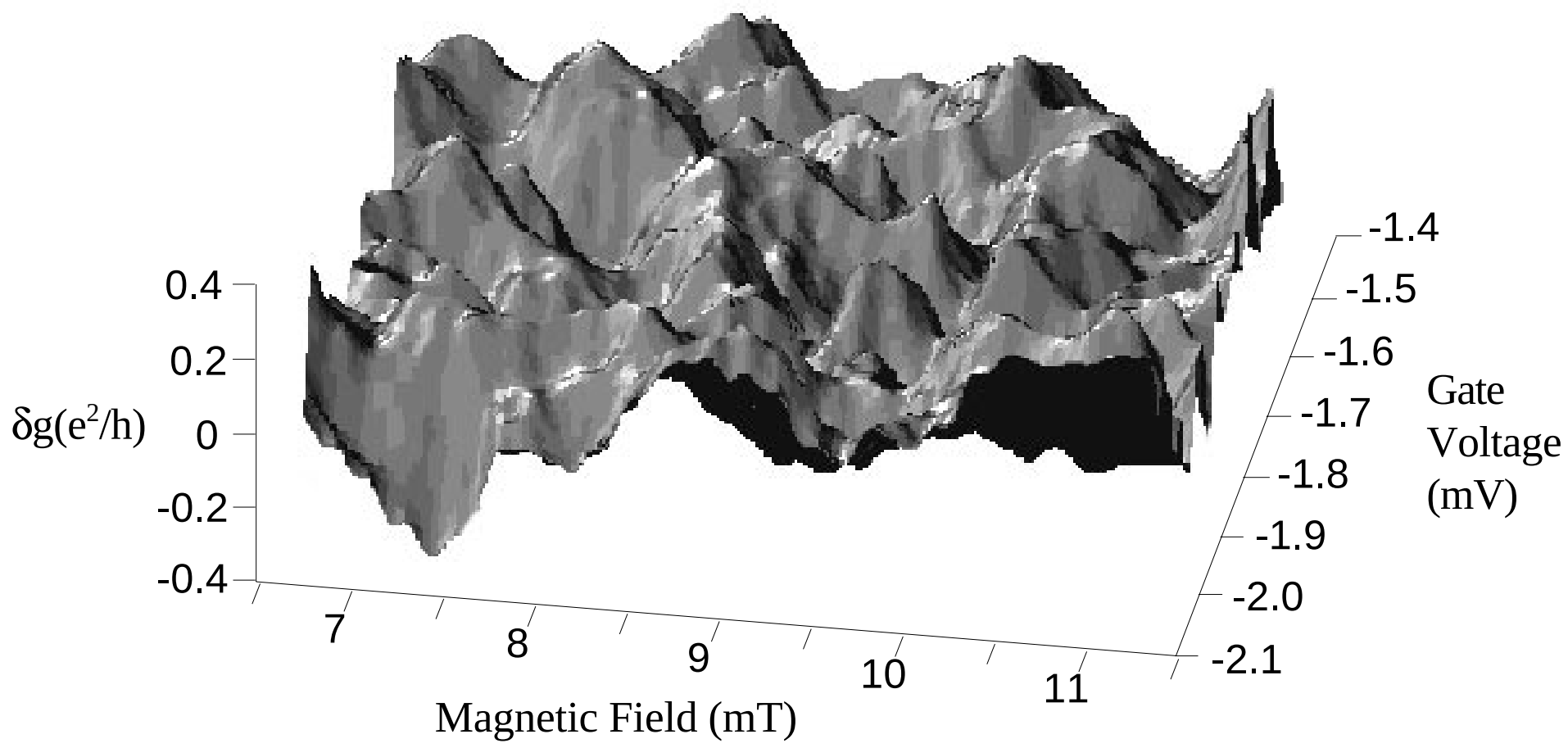

**Fig. 4.** Landscape of conductance fluctuations for the dot shown in the inset at ~ 100 mK. Both magnetic field and shape distorting gate voltage act as sources of conductance fluctuations with similar statistics and correlations.

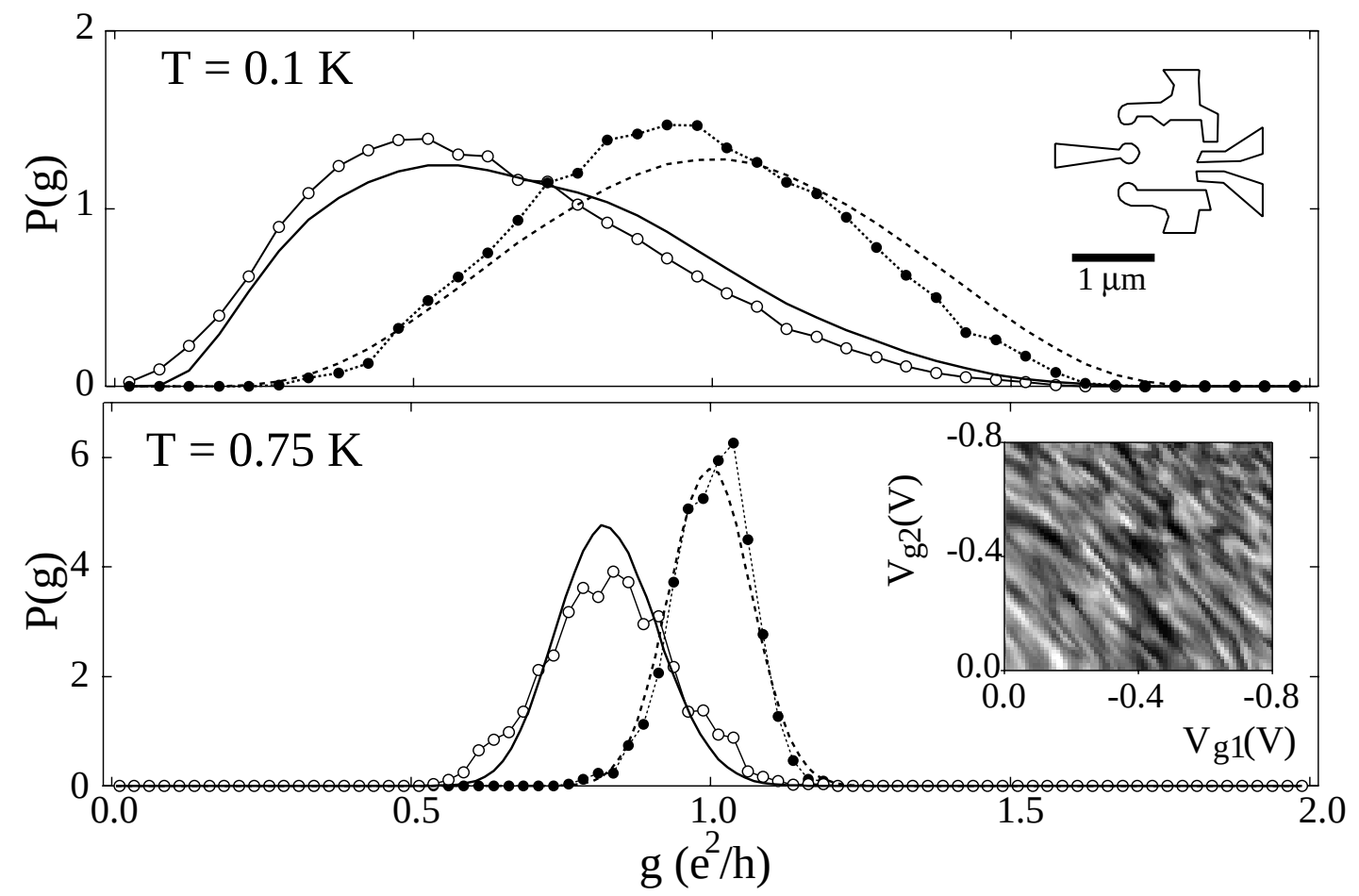

**Fig. 5.** Full distributions P(g) of conductance, at low and high temperature, showing excellent agreement between theory (curves) and experiment (markers). Distributions are for zero (open) and nonzero (solid) magnetic fields. Lower inset shows a 2D landscape of conductance as a function of voltages on shape distorting gates, used to gather statistics. Adapted from Huibers, *et al.*, Ref. 22.

We will not discuss conductance in the regime of Coulomb blockade except to note that the statistics[27,28] and parametric correlations[29] of conductance in this regime have also been investigated theoretically and again are found to be in good agreement with experiment.[30,31,32]

## 3. Feedback and Control

Quantum interference makes dots extremely sensitive to changes in external parameters. This can be used both to record changes in the environment, for instance as a magnetometer[16], or to actively feed signals back into the device. One would like to know how controllable such quantum system is and generally what are the rules for applying feedback along the lines of other recent work concerning the control of quantum systems.[5,6]

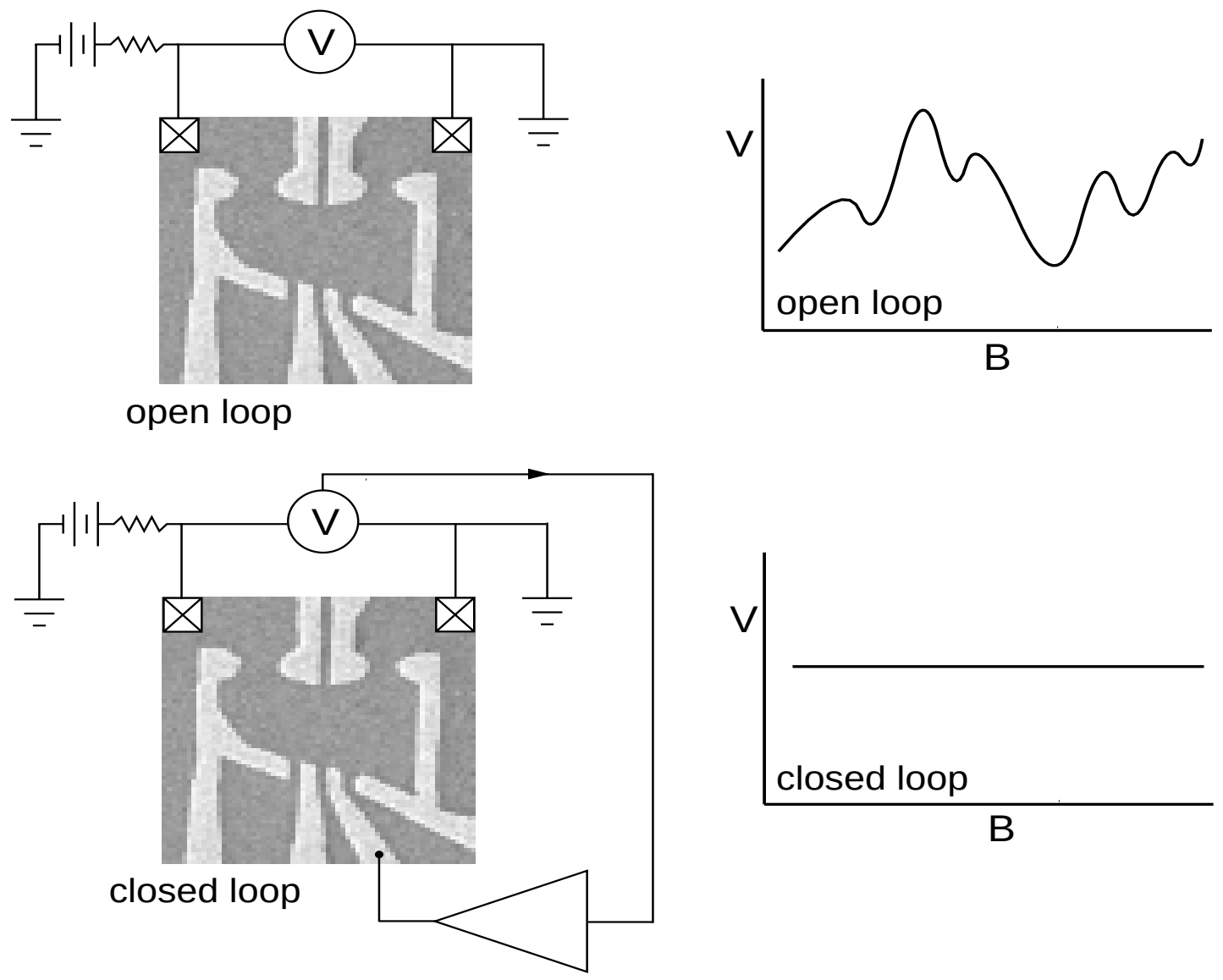

**Fig. 6.** An example of feedback control of mesoscopic fluctuations. In the top set-up, the dot runs open loop and shows conductance fluctuations as a function of magnetic field. Is it possible to eliminate fluctuations using feedback to one of the shape distorting gates?

We consider a simple example of feedback control illustrated schematically in Fig. 6. In an open-loop configuration, a quantum dot run in current bias mode shows large, random voltage fluctuations as an external perpendicular magnetic field is changed (Fig. 6(a)). One could imagine such fluctuations arising from a number of other sources as well, for instance from fluctuating dopant charge states that change the confining potential as a func-tion of time.[33] The question we ask is, can one *eliminate* conductance fluctuations using feedback, as suggested in Fig. 6(b)? For small signals (e.g. $B < B_c$ in Fig. 6), linear feedback is effective, though initially one does not know the proper sign of the feedback. To feedback-control larger signals, one must face the problem of controlling a random signal with a random signal.

By considering classical feedback, i.e. with its source external to the system and not coherent, the question is reduced to considering the problem of navigating along level contours (lines of constant conductance) in the ($m$ + 1)-dimensional space of the independent external parameter plus $m$ control parameters. Figure 6(b) illustrates the case of a single feedback signal applied to a gate voltage, so that the space being navigated is two dimensional and the level contours generically formed closed paths, as shown in Fig. 7. Figure 7 is actually based on experimental data for two shape distorting gates rather than one gate and magnetic field. However, the *statistical* properties of these contour lines are insensitive to which external parameters are being varied. To understand the range over which feedback signals can be used eliminate mesoscopic fluctuations, we turn to a statistical picture of contour lines of an ($m$ + 1)-dimensional random landscape with statistics and correlations as given above.

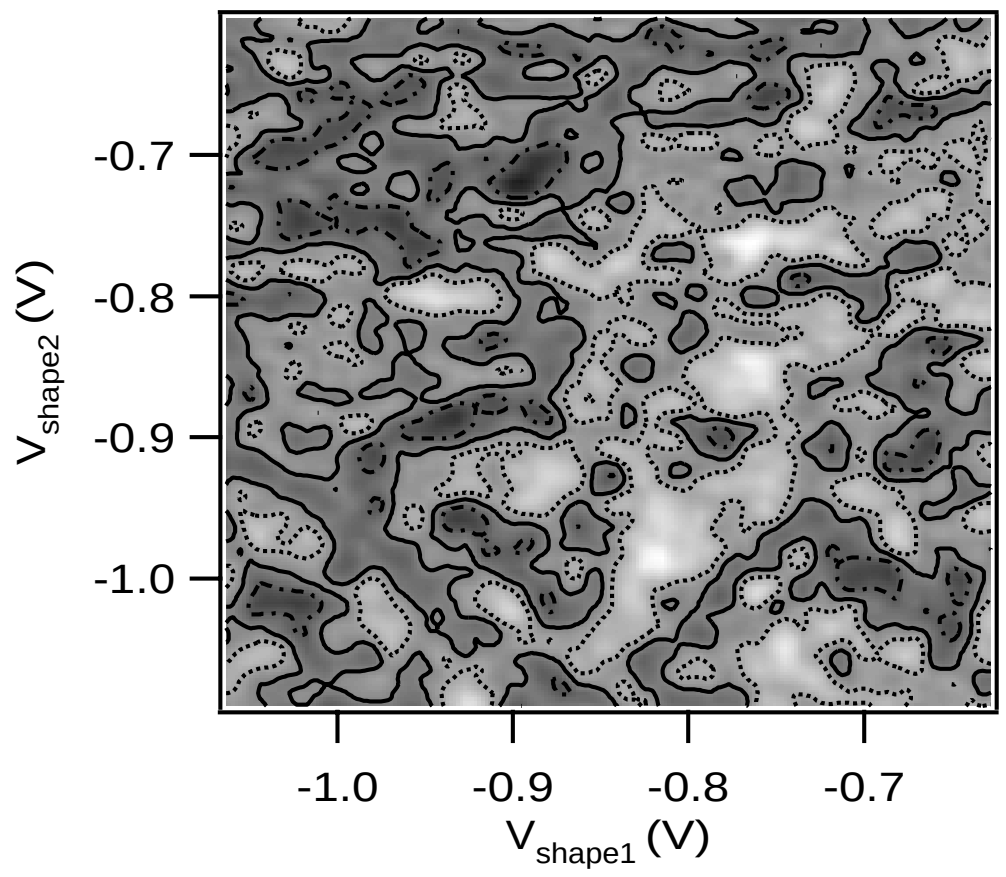

**Fig. 7.** Contour lines of conductance on an experimental landscape in the 2D space of voltages on shape-distorting gates. Contours at low (dashed) and high (dotted) conductances form small loops, while the conductance near the average (solid) percolates across the landscape. The size of a contour loop indicates the range of controllability.

## 4. Feedback and Statistical Topography

By restating the condition for the controllability of a mesoscopic device with $m$ controls in terms of the problem of navigating a contour line in an ($m$ + 1)-dimensional random landscape, we are able to draw on known results from continuum percolation the-ory[14,34,35] concerning the range over which such navigation is likely. An illustration of continuum percolation in two dimensions is shown in Fig. 8. White areas represent regions of the landscape above a given contour; black regions represent regions below that contour. Figures 8(a) and (c) show that in two dimensions, any generic contour line (the border between black and white) will only form a small loop with a size comparable to the characteristic spatial scale of the fluctuations. Only one contour, which, for the case of a symmetrically random landscape is the average of the function, will “percolate” i.e. traverse the landscape (Fig. 8(b)). Away from this point, the typical diameter of closed contours—i.e. the range of controllability—decreases like $|h|^{-4/3}$, where $h$ is the height (positive or negative) of the contour away from the percolating contour.[36] We conclude that with a single control parameter, one is typically only able to control the conductance over a range comparable to the characteristic fluctuation scale of the independent parameter except very close to the percolating contour. This is not a very desirable situation.

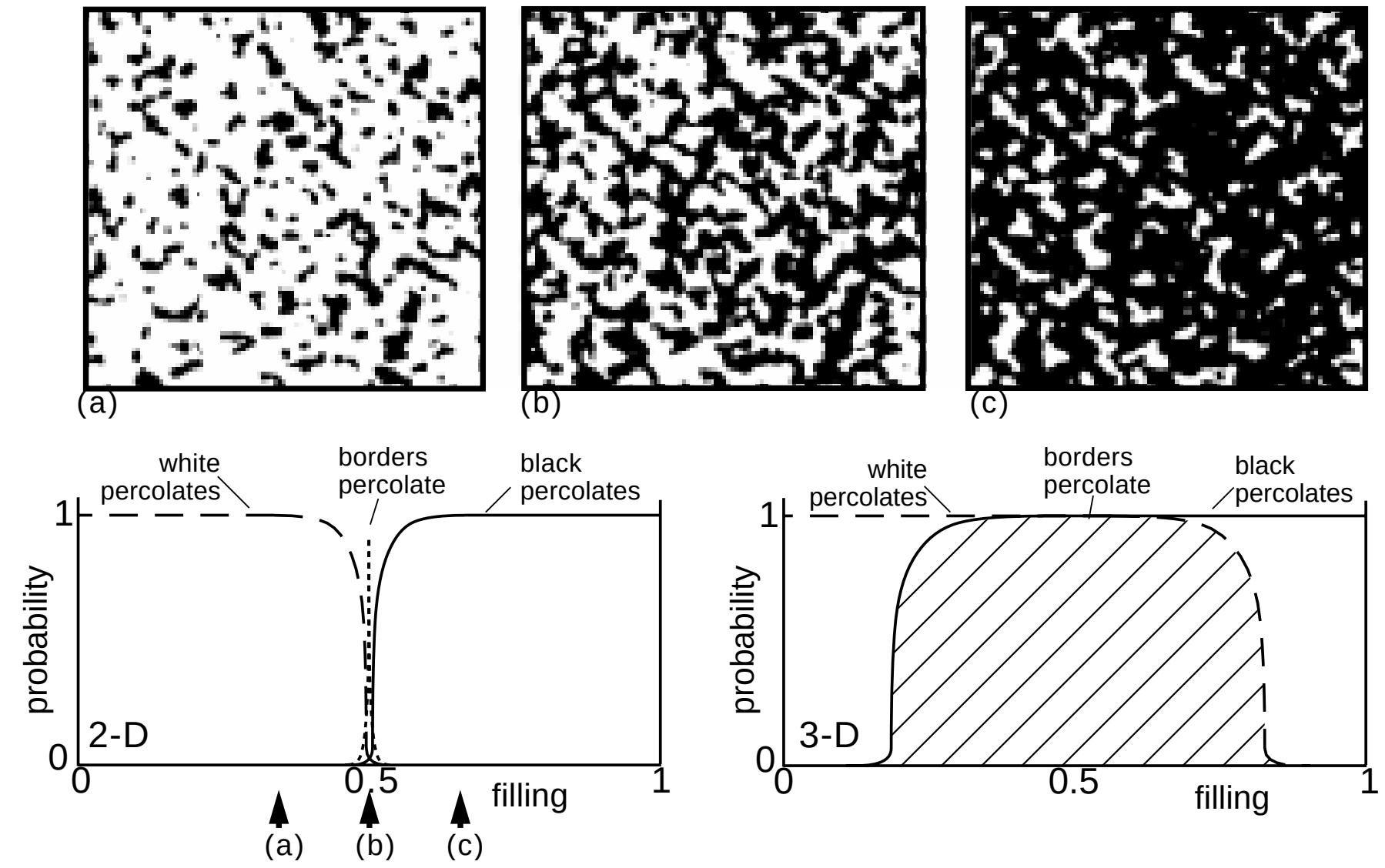

**Fig. 8.** (a,b,c) Regions above (white) and below (black) a particular contour line of a 2D random surface as an example of continuum percolation. In (a) and (c) the contour (i.e. the border between white and black) forms small closed loops. In (b) the contour percolates across the landscape. Percolation only occurs for the contour at the average height of the random surface. Below, left: the three contours are indicated. Below, right: Continuum percolation in 3D, indicating a finite range of heights where contours percolate. This implies that two control parameters suffice to use feedback to eliminate conductance fluctuations, as explained in the text.

The situation is quite different when $(m + 1) = 3$, i.e. with two control parameters. In this case, there exists a range of contour heights for which contours percolate, as illustrated in the lower-right panel of Fig. 8.[14,34] We therefore conclude that in order to control mesoscopic fluctuations over a broad range of conductances, two control parameters, perhaps two shape-distorting gates, are both necessary and sufficient. Further broadening of the range of percolating contours can be realized using more controls ($m > 2$), but in this case exactly how to employ the feedback becomes a difficult problem.

## 5. Conclusions

This paper presented a simple example in which the statistical character of fluctuations of coherent electron transport, ultimately arising from quantum chaos, can be used to derive new bounds on the controllability of this system when used in a feedback mode. Quantum dots are an ideal technology in which to explore feedback and coherent control in quantum systems, simply because changes to the system can be realized so easily using electrostatic gates. Incorporating time-domain aspects of coherent control, as currently done in quantum chemistry[5] is the logical next step for this application. Work continues in this direction.

## 6. Acknowledgements

We thank numerous experimental colleagues at Stanford, Robert Westervelt at Harvard, Ken Campman , Art Gossard (UCSB), Cem Duruöz, and James Harris (Stanford). We acknowledge support from the Army Research Office, the Office of Naval Research, the NSF-NYI and PECASE programs.